\documentclass[12pt]{article}%
\usepackage{amsmath}
\usepackage{graphicx}
\usepackage{amsfonts}
\usepackage{amssymb}%
\providecommand{\U}[1]{\protect\rule{.1in}{.1in}}
\providecommand{\U}[1]{\protect\rule{.1in}{.1in}}
\begin{document}

\title{A Compact Approximate Solution to the Friedel-Anderson Impuriy Problem}
\author{Gerd Bergmann\\Department of Physics\\University of Southern California\\Los Angeles, California 90089-0484\\e-mail: bergmann@usc.edu}
\date{\today}
\maketitle

\begin{abstract}
An approximate groundstate of the Anderson-Friedel impurity problem is
presented in a very compact form. It requires solely the optimization of two
localized electron states and consists of four Slater states (Slater
determinants). The resulting singlet ground state energy lies far below the
Anderson mean field solution and agrees well with the numerical results by
Gunnarsson and Schoenhammer, who used an extensive $1/N_{f}$-expansion for a
spin 1/2 impurity with double occupancy of the impurity level.


\newpage

\end{abstract}

\section{Introduction}

The properties of magnetic impurities in a metal is one of the most
intensively studied problems in solid state physics. Although some of the
experimental anomalies were already discovered in the 1930's it is still a
subject of great interest. The work of Friedel \cite{F28} and Anderson
\cite{A31} laid the foundation to understand why some transition metal
impurities form a magnetic local moment while others don't. They considered a
host with an s-band in which a transition metal atom is dissolved. The
s-electrons can hop onto the d-impurity via the hopping matrix element
$V_{sd}$. The ten-fold degeneracy of a real d-impurity is simplified and
reduced to a two-fold degeneracy for spin up and spin down. If both states are
occupied they repel each other due to the Coulomb exchange energy. This yields
the Friedel-Anderson Hamiltonian
\begin{equation}
H_{FA}=\sum_{\sigma}\{\sum_{\nu=1}^{N}\varepsilon_{\nu}c_{\nu\sigma}^{\ast
}c_{\nu\sigma}+E_{d}d_{\sigma}^{\ast}d_{\sigma}+\sum_{\nu=1}^{N}V_{sd}%
(\nu)[d_{\sigma}^{\ast}c_{\nu\sigma}+c_{\nu\sigma}^{\ast}d_{\sigma}%
]\}+Un_{d+}n_{d-} \label{hfa0}%
\end{equation}
Here a finite s-band with $N$ states is used. The $c_{\nu\sigma}^{\ast}$ and
the $d_{\sigma}^{\ast}$ are the creation operators of the (free) s-electrons
and the d-impurity. The $d_{\sigma}$-states are assumed to be orthogonal to
the s-states $c_{\nu}^{\ast}$.

Kondo \cite{K8} showed that multiple scattering of conduction electrons by a
magnetic impurity yields a divergent contribution to the resistance in
perturbation theory. In the following three decades a large number of
sophisticated methods were applied to better understand and solve the Kondo
and Friedel-Anderson model, and it was shown that at zero temperature a
Friedel-Anderson impurity is in a non-magnetic state. To name a few of these
methods: scaling \cite{A51}, renormalization \cite{W18}, \cite{F30},
Fermi-liquid theory \cite{N14}, \cite{N5}, slave-bosons (see for example
\cite{N7}), large-spin limit \cite{G19}, \cite{B103}, and the Bethe-ansatz
\cite{W12}, \cite{S29}. For a review see \cite{H20}. After decades of research
exact solutions of the Kondo and Friedel-Anderson problems were derived
\cite{W12}, \cite{A50} representing a magnificent theoretical achievement.

The exact solution does not solve all questions. It uses a s-electron band
with a constant density of states which extends from minus infinity to plus
infinity (the cut-off is only performed at the end of the calculation).
Furthermore it is such a complex solution that only a limited number of
parameters can be calculated and many non-critical or non-divergent
contributions are neglected. For the majority of practical problems one uses
approximate solutions. One particularly popular method is the large-spin
method which will be discussed below.

While the single impurity problem is intensively studied and well understood
the many-impurity problem and the periodic Anderson problem are still in a
rather incomplete state \cite{Z11}. Any simplified treatment of the single
impurity may provide a new tool to improve the treatment of the latter.

In this paper I wish to introduce a new method to treat the Friedel-Anderson
problem. This approach can be best explained by a discussion of the ground
state of the Friedel Hamiltonian. This is done in chapter 2. In chapter 3 the
singlet state of the Friedel-Anderson model is derived. In chapter 4 the
results are discussed. In the appendix some details of the calculation are summarized.

\section{ The Friedel Resonance}

The author's method to treat the Friedel-Anderson problem can be best
explained by a discussion of the ground state of the Friedel (resonance)
Hamiltonian $H_{Fr}$. For an s-band with N states and a $d$ resonance $H_{Fr}$
has the form:%

\begin{equation}
H_{Fr}=\sum_{\nu=1}^{N}\varepsilon_{\nu}c_{\nu}^{\ast}c_{\nu}+E_{d}d^{\ast
}d+\sum_{\nu=1}^{N}V_{sd}(\nu)[d^{\ast}c_{\nu}+c_{\nu}^{\ast}d] \label{hfr}%
\end{equation}
Since $H_{Fr}$ is identical for spin up and down, I will ignore the spin at
the moment.

As shown in ref. \cite{B91}, \cite{B92} the exact ground state of $H_{Fr}$
with $n$ (spinless) electrons can be written in the form%

\begin{equation}
\Psi_{Fr}=\left[  A^{\prime}a_{0}^{\ast}+B^{\prime}d^{\ast}\right]
\prod_{i=1}^{n-1}a_{i}^{\ast}\Phi_{0} \label{yfr}%
\end{equation}

Here $\Phi_{0}$ is the vacuum state and $a_{0}^{*}$ is a localized state which
is built from the states of the s-band%

\begin{equation}
a_{0}^{*}=\sum_{\nu=1}^{N}\alpha_{\nu}^{0}c_{\nu}^{*} \label{afr}%
\end{equation}

The $a_{i}^{*}$ ($1\leq i\leq N-1$) together with $a_{0}^{*}$ represent a new
basis. The $a_{i}^{*}$ are orthogonal to $a_{0}^{*}$ and to each other and
their $(N-1)$ sub-matrix of the s-band Hamiltonian $H_{0}=\sum\varepsilon
_{\nu}n_{\nu}$ is diagonal. (The construction of the states $\{a_{0}^{*}%
,a_{i}^{*}\}$ is discussed in appendix A). The states $a_{i}^{*}$ are uniquely
determined from the state $a_{0}^{*}$. Their form is
\begin{equation}
a_{i}^{*}=\sum_{\nu=1}^{N}\alpha_{\nu}^{i}c_{\nu}^{*}%
\end{equation}

In this new basis the (spin independent) Friedel Hamiltonian can be written as%

\begin{align}
H_{Fr}  &  =\sum_{i=1}^{N-1}E\left(  i\right)  a_{i}^{\ast}a_{i}+E\left(
0\right)  a_{0}^{\ast}a_{0}+\sum_{i=1}^{N-1}V_{fr}^{a}\left(  i\right)
\left[  a_{0}^{\ast}a_{i}+a_{i}^{\ast}a_{0}\right] \label{hfr'}\\
&  +E_{d}d^{\ast}d+V_{sd}^{a}(0)[d^{\ast}a_{0}+a_{0}^{\ast}d]+\sum_{i=1}%
^{N-1}V_{sd}^{a}\left(  i\right)  \left[  d^{\ast}a_{i}+a_{i}^{\ast}d\right]
\nonumber
\end{align}
where%

\begin{equation}%
\begin{array}
[c]{ccc}%
E\left(  i\right)  =\sum_{\nu}\alpha_{\nu}^{i}\varepsilon_{\nu}\alpha_{\nu
}^{i} &  & E\left(  0\right)  =\sum_{\nu}\alpha_{\nu}^{0}\varepsilon_{\nu
}\alpha_{\nu}^{0}\\
&  & \\
V_{sd}^{a}\left(  i\right)  =\sum_{\nu}V_{sd}\left(  \nu\right)  \alpha_{\nu
}^{i} &  & V_{fr}^{a}\left(  i\right)  =\sum_{\nu}\alpha_{\nu}^{i}%
\varepsilon_{\nu}\alpha_{\nu}^{0}\\
&  &
\end{array}
\label{efr}%
\end{equation}

In the hamiltonian (\ref{hfr'}) the first three terms represent the free
electron hamiltonian. The $a_{0}^{\ast}$-state represents an artificial
Friedel resonance state. (AFR state). It is interesting to note that the
$d^{\ast}$-state and the localized $a_{0}^{\ast}$-state in (\ref{hfr'}) are on
equal footing. The AFR state $a_{0}^{\ast}$ is a sister state to the state
$d^{\ast}.$

The terms with the matrix elements $V_{fr}^{a}\left(  i\right)  $ and
$V_{sd}^{a}\left(  i\right)  $ yields the hopping between $a_{i}^{\ast}$ and
$d^{\ast}$ and $a_{i}^{\ast}$ and $a_{0}^{\ast}$. For the state $\left(
A^{\prime}a_{0}^{\ast}+B^{\prime}d^{\ast}\right)  $ the individual hopping
matrix elements cancel each other, making $\Psi_{Fr}$ the ground-state.

Now I consider a conduction band and a $d$ state with two spin components.
Again I take the Friedel Hamiltonian as given by eq. (\ref{hfr}). Then the
ground state is given by the product of the spin up and down states of eq.
(\ref{yfr}). This exact ground state can be written as%

\begin{align}
\Psi_{MS}  &  =\left[  A_{-}a_{0-}^{\ast}+B_{-}d_{-}^{\ast}\right]  \left[
A_{+}a_{0+}^{\ast}+B_{+}d_{+}^{\ast}\right]  \prod_{\sigma,i=1}^{n-1}%
a_{i\sigma}^{\ast}\Phi_{0}\nonumber\\
\  &  =\left[  Aa_{0-}^{\ast}a_{0+}^{\ast}+Bd_{-}^{\ast}a_{0+}^{\ast}%
+Ca_{0-}^{\ast}d_{+}^{\ast}+Dd_{-}^{\ast}d_{+}^{\ast}\right]  \prod
_{\sigma,i=1}^{n-1}a_{i\sigma}^{\ast}\Phi_{0}\label{y0}\\
&  =A\Psi_{A}+B\Psi_{B}+C\Psi_{C}+D\Psi_{D}\nonumber
\end{align}

where%

\begin{equation}%
\begin{array}
[c]{ccccccc}%
A_{+}^{2}+B_{+}^{2} & = & 1 & , & A_{-}^{2}+B_{-}^{2} & = & 1\\
A & = & A_{+}A_{-} & , & B & = & A_{-}B_{+}\\
C & = & A_{+}B_{-} & , & D & = & B_{+}B_{-\,}%
\end{array}
\label{no0}%
\end{equation}
Each of the four states $\Psi_{A}$, $\Psi_{B}$, $\Psi_{C}$ and $\Psi_{D}$ is
normalized and they are all orthogonal to each other because they differ in
the occupations of the $d_{+}^{\ast}$ or $d_{-}^{\ast}$-state. In Fig.1 the
four components $\Psi_{A}$, $\Psi_{B}$, $\Psi_{C}$ and $\Psi_{D}$ of the state
$\Psi_{MS}$ are graphically shown.

In the present case one has $A_{+}=A_{-}$ and $B_{+}=B_{-}$. With the right
choice of $a_{0}^{\ast}$ the above state is the exact ground state of the
Friedel Hamiltonian.%

\[%
{\includegraphics[
height=3.6222in,
width=4.1752in
]%
{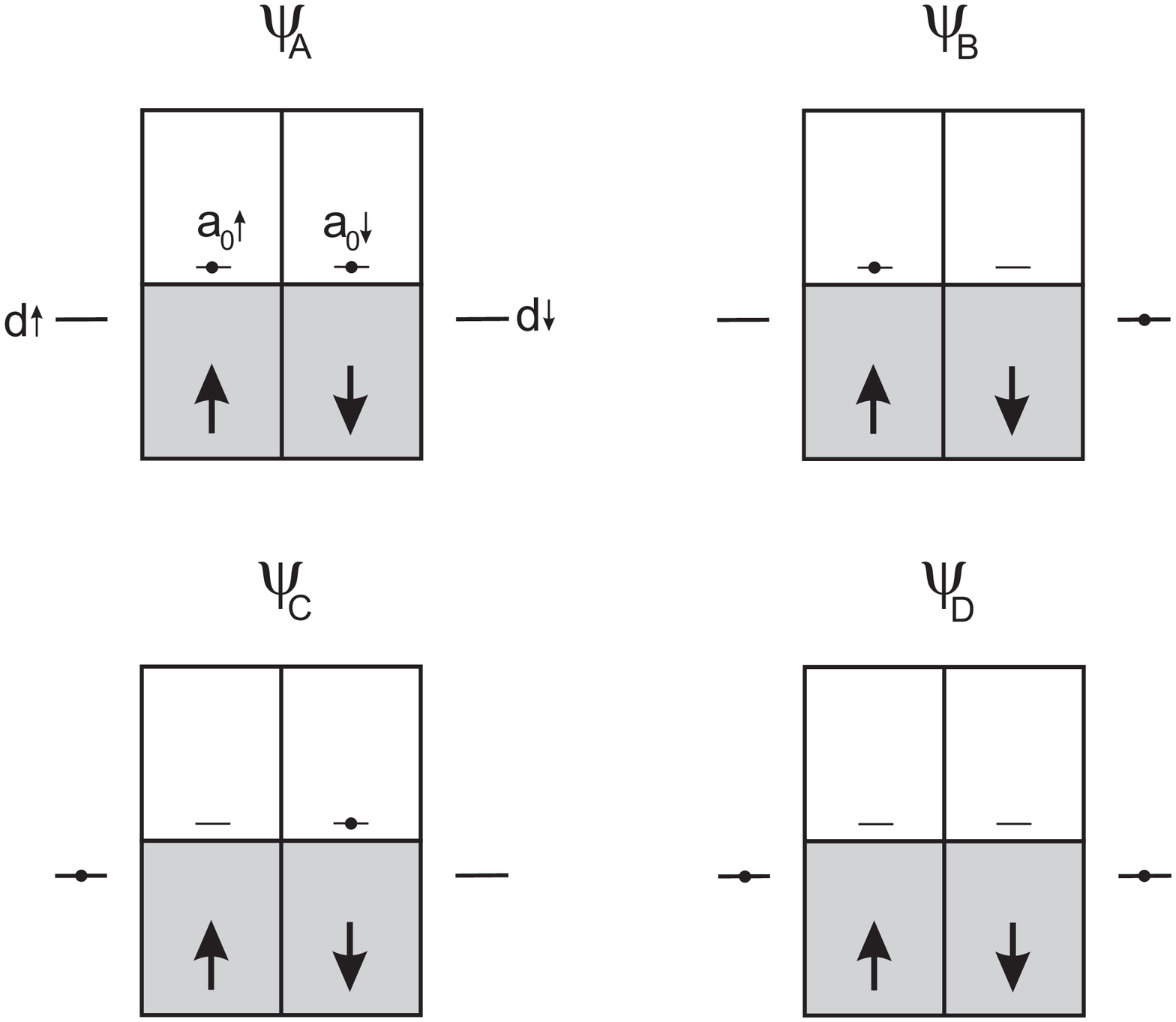}%
}%
\]
Fig.1: The four Slater states used in the present calculation. In each
component either the $d^{\ast}$-state or the AFR state $a_{0}^{\ast}$ is
occupied.\newline

\section{ The Friedel-Anderson Model}

In the presence of a Coulomb exchange interaction $U$ one can use the general
form of the state $\Psi_{MS}$ as an approximate ground state. This state
consists of four Slater states and possesses considerable flexibility. One may
drop the condition $A_{+}=A_{-}$ and $B_{+}=B_{-}$ and use $A,B,C,D$ as free
parameters fulfilling the condition%
\[
\left\vert A\right\vert ^{2}+\left\vert B\right\vert ^{2}+\left\vert
C\right\vert ^{2}+\left\vert D\right\vert ^{2}=1
\]
and optimizes these coefficients. Far more important is the optimization of
the states $a_{0+}$ and $a_{0-}$. With $a_{0+},$ $a_{0-}$ the full new bases
$\left\{  a_{i+}^{\ast}\right\}  ,\left\{  a_{i-}^{\ast}\right\}  $ are
uniquely determined. Details of the numerical optimization are discussed in
the appendix.

In the following calculations a half filled s-subband is used, i.e. the number
of occupied states in each s-subband is $n=N/2.$ The energy expectation value
of the state $\Psi_{MS}$ is calculated. For zero Coulomb exchange energy $U$
\ the spin up and down bases are identical as are the coefficients $B$ and
$C$. With increasing $U$ the spin up and down bases are shifted with respect
to each other, the coefficients $B$ and $C$ differ and a magnetic moment
develops. The size of this moment and a comparison with the mean field
solution is discussed elsewhere \cite{B152}.

\subsection{The singlet state}

The ground state of the Friedel-Anderson problem is a singlet state. From
$\Psi_{MS}$ one can construct a mirror state by exchanging spin up and down.
Combining the two states yields then a singlet state which I denote as
$\Psi_{SS}$. It is given by the following expression
\[
\Psi_{SS}=\Psi_{MS}\left(  \uparrow\downarrow\right)  \mp\Psi_{MS}\left(
\downarrow\uparrow\right)
\]%
\begin{align*}
&  =\left[  Aa_{0-\downarrow}^{\ast}a_{0+\uparrow}^{\ast}+Bd_{-\downarrow
}^{\ast}a_{0+\uparrow}^{\ast}+Ca_{0-\downarrow}^{\ast}d_{+\uparrow}^{\ast
}+Dd_{-\downarrow}^{\ast}d_{+\uparrow}^{\ast}\right]  \prod_{i=1}%
^{n-1}a_{i+\uparrow}^{\ast}\prod_{i=1}^{n-1}a_{i-\downarrow}^{\ast}\Phi_{0}\\
&  \mp\left[  Aa_{0-\uparrow}^{\ast}a_{0+\downarrow}^{\ast}+Bd_{-\uparrow
}^{\ast}a_{0+\downarrow}^{\ast}+Ca_{0-\uparrow}^{\ast}d_{+\downarrow}^{\ast
}+Dd_{-\uparrow}^{\ast}d_{+\downarrow}^{\ast}\right]  \prod_{i=1}%
^{n-1}a_{i+\downarrow}^{\ast}\prod_{i=1}^{n-1}a_{i-\uparrow}^{\ast}\Phi_{0}%
\end{align*}
The sign $\mp$ is chosen so that one obtains a singlet state. This state is
not normalized and the \textquotedblright$B$\textquotedblright\ and
\textquotedblright$C$\textquotedblright\ components are not orthogonal to each
other. This introduces some additional terms in the ground-state energy.
Furthermore the matrix elements between the states $\Psi_{MS\uparrow}$ and
$\Psi_{MS\downarrow}$ become determinants of single electron matrix elements.
This is discussed in the appendix.

For the numerical calculation an s-band with $N$ electron states $c_{\nu
}^{\ast}$ is used (in the following I denote single electron states by their
creation operator). A logarithmic energy scale is used, as introduced by
Wilson \cite{W18} in his Kondo paper. One uses a finer and finer energy scale
when approaching the Fermi energy $\varepsilon_{F}=0$. A brief description of
these electron states is given in the appendix.

The ground-state energy of the singlet state is shown in Fig.3 as a function
of $U$ with $E_{d}=-U/2$ and compared with the mean field ground state energy.
Its energy clearly lies below the energy of the mean field state%

\[%
{\includegraphics[
height=3.1756in,
width=3.8232in
]%
{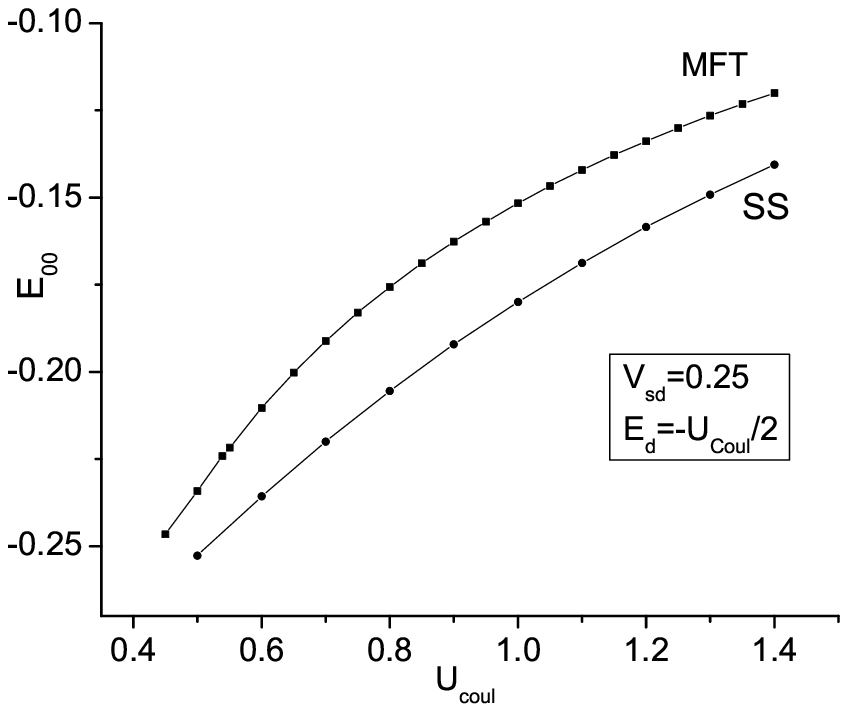}%
}%
\]
Fig.2: A comparison between the ground-state energies of the Anderson's mean
field calculation and the singlet state $\Psi_{ss}.$%
\[
\]

\section{Discussion}

\subsection{Comparison with the large $N_{f}$-expansion}

A number of approximate solutions have been suggested in the literature in
which a localized electron state forms a singlet state with the magnetic
impurity, see for example \cite{Y5}, \cite{A49},\cite{G19}. They have been
suggested for the Kondo problem and the Friedel-Anderson model. One
particularly popular approximation is the large $N_{f}$ expansion (see for
example \cite{G19}, \cite{B103}). In the large $N_{f}$ expansion one assumes
that the impurity has a large total angular momentum $J_{f}$ ($J_{f}$ because
this method is often used for f-impurities). The \textquotedblright
spin\textquotedblright\ has then a degeneracy of $N_{f}=(2J_{f}+1)$ of the
total angular momentum states. In the limit of infinite $N_{f}$ (the large
$N_{f}$ limit) this method yields an exact ground state. For smaller spin, in
particular for spin $1/2,$ one generally performs an expansion in powers of
$1/N_{f}.$

Gunnarsson and Schoenhammer (G\&S) \cite{G34} applied the large $N_{f}$ method
to a finite Coulomb interaction and spin $\dfrac{1}{2}$, including double
occupancy of the impurity level. They calculated the ground-state energy in
different approximations. G\&S give their energy parameters in units of
$\left[  \text{eV}\right]  .$ I denote their parameters with the index
\textquotedblright GS\textquotedblright. These parameters are: half the
bandwidth $B_{GS}$, the d-state energy $E_{d,GS}$, the Coulomb energy $U_{GS}%
$. For the s-d-hopping transition they use an elliptic form
\begin{equation}
\left[  V\left(  \varepsilon\right)  \right]  ^{2}g\left(  \varepsilon\right)
=\dfrac{2V_{GS}^{2}}{\pi B_{GS}^{2}}\sqrt{\left(  B_{GS}^{2}-\varepsilon
^{2}\right)  } \label{V_GS}%
\end{equation}
where $g\left(  \varepsilon\right)  $ is the density of states (per spin). All
these parameters are energies or potentials. By dividing these energy
parameter by $B_{GS}$ one obtains the appropriate parameters for the present
calculation. When the numerical calculation is completed the resulting
ground-state energy must be multiplied with $B_{GS}$ for a comparison with
GS's results.

Gunnarsson and Schoenhammer calculated in ref. \cite{G34} the ground-state
energy (for $N_{f}=2$) with the following parameters: $B_{GS}=6$ eV,
$U_{BS}=5$ eV, $E_{d,BS}=-2.5$ eV. They performed two calculations, one for
$V_{GS}=1$eV and another for $V_{GS}=2$eV. The results are shown in tables
\ref{T4} and \ref{T5}. The first column gives the electron states used in the
calculation (for details see \cite{G34} and the appendix). The second column
gives the calculated ground-state energies. In the third, fourth and fifth
columns the symbols $f^{0},f^{1},f^{2}$ give the probabilities for a
d-occupation of 0,1,2. The last column gives the power of the $\left(
1/N_{f}\right)  $-expansion. The last row gives the results of the present
calculation for the singlet state. The values for $f_{0},f_{1},f_{2}$ agree
perfectly. Also the ground-state energies are quite close with $E_{GS}%
=-0.245eV$ \ \ and $E_{0}=-0.239eV$ of the present calculation.%

\begin{table}[tbp] \centering
\begin{tabular}
[c]{|l|l|l|l|l|l|}\hline
\textbf{states} & $E_{0}\left[  \text{eV}\right]  $ & $f_{0}$ & $f_{1}$ &
$f_{2}$ & \textbf{param.}\\\hline%
$\vert$%
0$\rangle$+a+b & -0.108 & 0.001 & 0.974 & 0.025 & $\left(  1/N_{f}\right)
^{0}$\\\hline
+c+d+e & -0.238 & 0.031 & 0.938 & 0.031 & $\left(  1/N_{f}\right)  ^{1}%
$\\\hline
+f+g & -0.245 & 0.034 & 0.931 & 0.034 & $\left(  1/N_{f}\right)  ^{2}$\\\hline%
\begin{tabular}
[c]{l}%
singlet\\
state
\end{tabular}
& -0.239 & 0.035 & 0.931 & 0.034 & \\\hline
\end{tabular}%
\caption{
A comparision between the numerical results by Gunnarsson and
Schoenhammer and the author for the case of $N_{f}=2$. The parameters,
given in the units used by GS, are $B_{GS}=6$ eV, $E_{d,GS}=-2.5$ eV,
$V_{GS}=1.$ eV, $U_{GS}=5$ eV.
The first column gives the states included in the large spin-method, the second column
gives the groundstate energy. The third, fourth and fifth columns give the weight of zero,
single and double occupation of the d-states.
The sixth column gives the number of optimized
parameters (amplitudes) in this calculation.
\label{T4}}%
\end{table}%
\[
\]

For $V_{GS}=2eV$ the ground-state energy of the present calculation lies even
below the value of the $1/N_{f}$-expansion, as shown in table \ref{T5}.%

\begin{table}[tbp] \centering
\begin{tabular}
[c]{|l|l|l|l|l|l|}\hline
\textbf{states} & $E_{0}\left[  \text{eV}\right]  $ & $f_{0}$ & $f_{1}$ &
$f_{2}$ & \textbf{expans.}\\\hline%
$\vert$%
0$\rangle$+a+b & -0.628 & 0.141 & 0.778 & 0.081 & $\left(  1/N_{f}\right)
^{0}$\\\hline
+c+d+e & -1.126 & 0.140 & 0.745 & 0.115 & $\left(  1/N_{f}\right)  ^{1}%
$\\\hline
+f+g & -1.217 & 0.137 & 0.732 & 0.132 & $\left(  1/N_{f}\right)  ^{2}$\\\hline%
\begin{tabular}
[c]{l}%
singlet\\
state
\end{tabular}
& -1.234 & 0.140 & 0.722 & 0.138 & \\\hline
\end{tabular}%
\caption{
A comparision between the numerical results by Gunnarsson and
Schoenhammer and the author for the case of $V_{GS}=2eV$. Everything
else is identical to table 1
\label{T5}}%
\end{table}%

The state \textquotedblright g\textquotedblright\ requires the variation of
more than $15^{6}$ , i.e. more than 10$^{7}$ amplitudes. In the present
calculation the singlet state requires the variation of 2$N$=60 amplitudes.
Keeping this in mind, the resulting ground state of the present calculation is
rather compact.

\subsection{Properties of the Artificial Resonance State}

The states $a_{0+}^{\ast}$ and $a_{0-}^{\ast}$ are of particular importance of
the present treatment of the Friedel-Anderson impurity. They determine the
rotation of the s-electron basis in Hilbert space and therefore the solution
of the problem. We analyze the composition of $a_{0\pm}^{\ast}$ in terms of
the original s-state energies $\varepsilon_{\nu}$. As discussed above,
$a_{0\pm}^{\ast}$ is composed of the original s-basis $c_{\nu}^{\ast}$ with
the amplitudes $\alpha_{\nu\pm}^{0}$
\[
a_{0\pm}^{\ast}=%
{\textstyle\sum_{\nu=1}^{N}}
\alpha_{\nu\pm}^{0}c_{\nu\pm}^{\ast}%
\]
In Fig.3a,b the coefficients $\alpha_{\nu+}^{0}$ and $\alpha_{\nu-}^{0}$of the
states $a_{0+}^{\ast}$and $a_{0-}^{\ast}$ are plotted for the parameters:
$U_{Cou}=1,$ $E_{d}=-0.5$, $\left\vert V_{sd}\right\vert ^{2}=0.1$ and the
number of s-states is $N=32$. One recognizes that the amplitudes at large
absolute energies is very different for spin up and down. They are almost
mirror images.
\[%
\begin{tabular}
[c]{ll}%
{\includegraphics[
height=2.4682in,
width=2.9581in
]%
{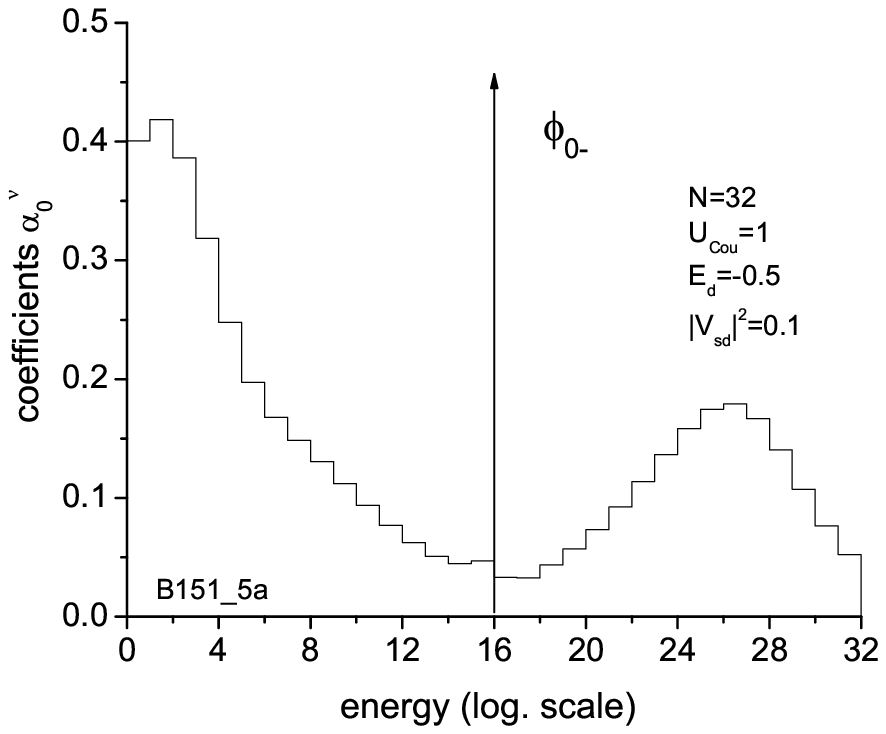}%
}%
&
{\includegraphics[
height=2.4724in,
width=2.9132in
]%
{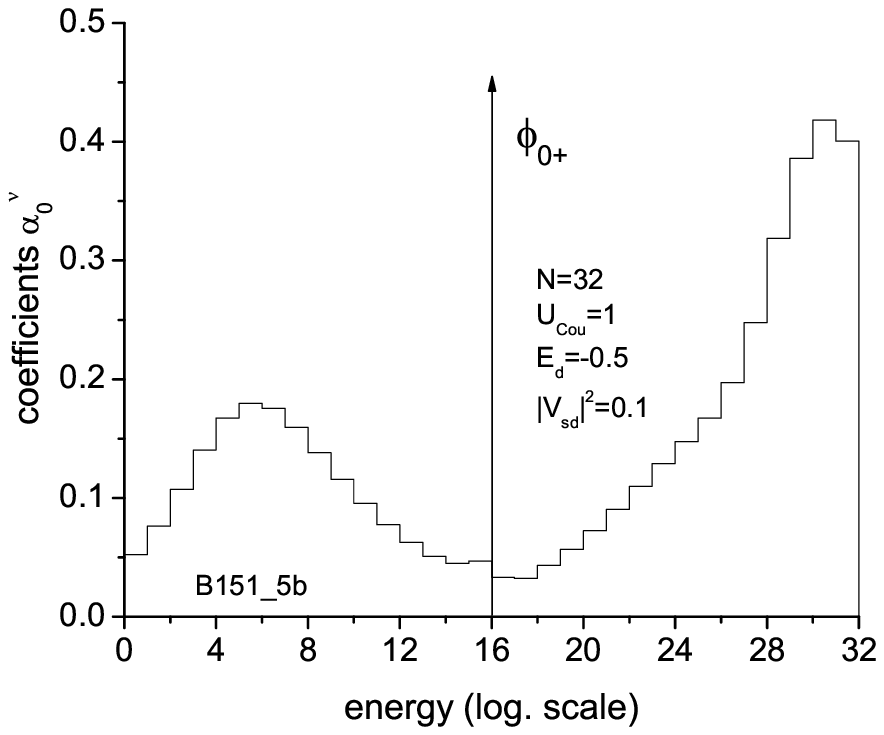}%
}%
\end{tabular}
\ \ \ \ \
\]
Fig.3a,b: The coefficients $\alpha_{0}^{\nu}$ for the AFR states $a_{0+}%
^{\ast}$ and $a_{0-}^{\ast}.$ The Wilson spectrum is used (for the region on
the left side of the arrow the numbers $\nu$ corresponds to an energy of
$E_{\nu}=-1/2^{\nu}$). The energy of the s-electron $c_{\nu}^{\ast}$ is
$\left(  E_{\nu}+E_{\nu-1}\right)  /2$. On the right side of the arrow one has
the corresponding positive energies.
\[
\]

For the analysis at small energies we plot the occupation density $\left\vert
\alpha_{0\pm}^{\nu}\right\vert ^{2}/\left(  E_{\nu}+E_{\nu-1}\right)  $ as a
function of $\nu$. In Fig.4a,b these densities are shown for $N=32$ and
$N=48.$ In the latter case the energy interval next to the arrow (zero energy)
is $1/2^{8}$ times smaller than for the left plot. Obviously the sub-division
at the Fermi energy is not yet sufficiently small at the left plot for $N=32$.%

\[%
\begin{tabular}
[c]{ll}%
{\includegraphics[
height=2.6094in,
width=3.0519in
]%
{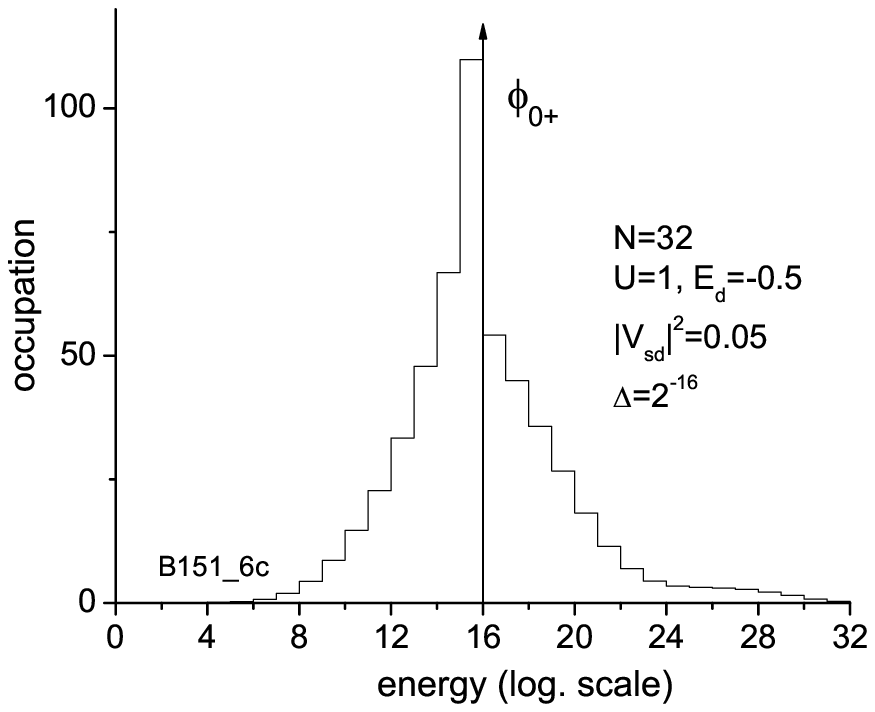}%
}%
&
{\includegraphics[
height=2.6119in,
width=3.0676in
]%
{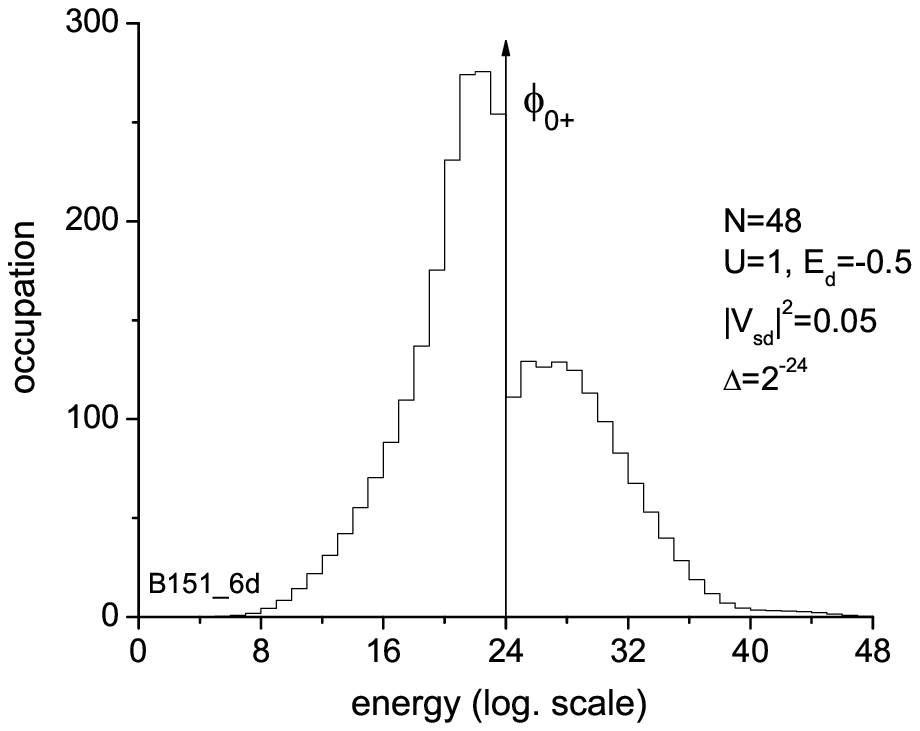}%
}%
\end{tabular}
\ \ \
\]
Fig.4a,b: The low energy occupation in the state $a_{0+}^{\ast}$ for different
sub-division of the energy close to the Fermi energy. In a) the smallest
sub-division is $\Delta=2^{-16}\thickapprox\allowbreak1.\,\allowbreak
5\times10^{-5}$ and in b) it is $\Delta=2^{-24}\thickapprox\allowbreak
6\times10^{-8}$. While in the left plot $\Delta$ is not yet small enough, one
observes in the right plot the occupation of $a_{0+}^{\ast}$ has saturated.%
\[
\]

While amplitudes and occupations for large energies were rather different for
$a_{0+}^{\ast}$ and $a_{0-}^{\ast}$ the occupation at small energies is almost
identical. This is shown in Fig.5 where the occupations of $a_{0+}^{\ast}$ and
$a_{0-}^{\ast}$ are plotted in the same figure. At energies close to the Fermi
energy the occupation of $a_{0+}^{\ast}$ and $a_{0-}^{\ast}$ are essentially
identical. On a linear energy scale at small energies the plots in Fig.4b and
Fig.5 are essentially identical.
\[%
\raisebox{-0pt}{\includegraphics[
height=3.3723in,
width=3.907in
]%
{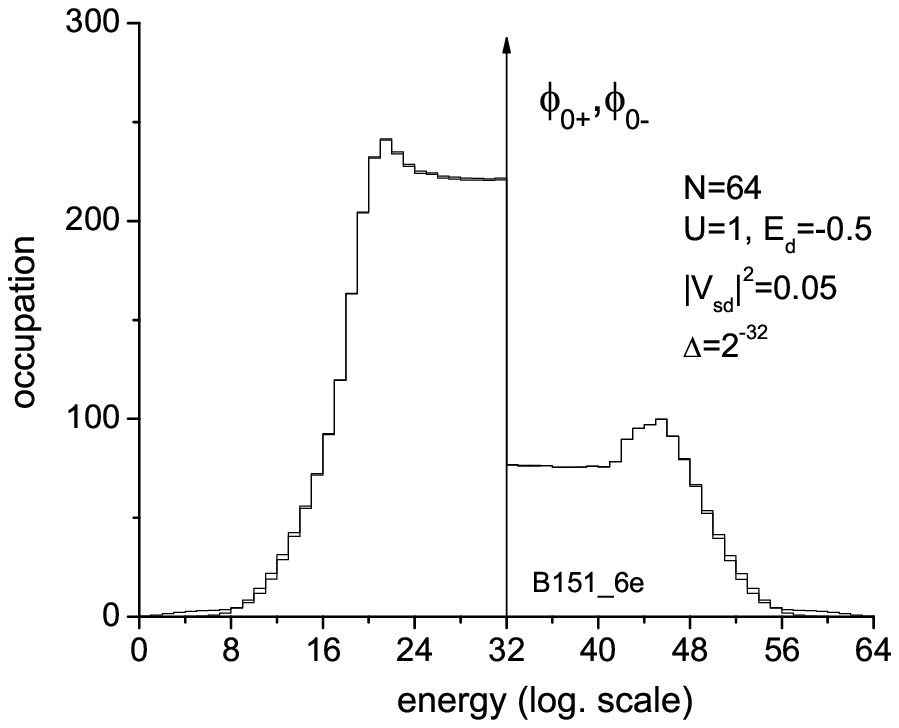}%
}%
\]
Fig.5: The occupation of the spin up and down AFR states for $N=52$. At small
energies (close to the center arrow) the two are almost identical. (The
difference between $\Phi_{0+}$ and $\Phi_{0-}$ is of the order of the width of
the curve.)
\[
\]

The average occupation density of the states $a_{0+}^{\ast}$ and $a_{0-}%
^{\ast}$ is 1/2 since the band ranges form -1 to +1. Therefore a density of
more than 100 is quite large.

The AFR states have weight at small and large energies. The weight at large
energies is responsible for the large "perturbative" part of the ground-state
energy. The weight at small energies is responsible for the anomalous behavior
at low temperatures, the Kondo effect.

\section{Conclusions}

In this paper I suggest a very compact approximate ground state for the
Friedel-Anderson impurity. Its center piece are two artificial resonance
states $a_{0+}^{\ast},a_{0-}^{\ast}$ for the spin up and down s-electrons.
These are combined with the d-electrons for spin up and down $d_{+}^{\ast
},d_{-}^{\ast}$ into two-electrons states of total spin zero, i.e. $\left[
Aa_{0-}^{\ast}a_{0+}^{\ast}+Bd_{-}^{\ast}a_{0+}^{\ast}+Ca_{0-}^{\ast}%
d_{+}^{\ast}+Dd_{-}^{\ast}d_{+}^{\ast}\right]  $. Then for each spin a new
s-electron basis $\left\{  a_{i\pm}^{\ast}\right\}  $ is built. These two
bases are completely determined by the AFR states. Finally the $\left(
n-1\right)  $ lowest states of the two basis are occupied yielding the
s-electron background $%
{\textstyle\prod\limits_{i=1,\sigma}^{n-1}}
a_{i\sigma}^{\ast}\Phi_{0}$. The compositions of the AFR states $a_{0+}^{\ast
},a_{0-}^{\ast}$ are calculated by numerical variation which rotates the
s-electron bases in Hilbert space. This ansatz is exact for a spin degeneracy
$N_{f}$ of ''1''\ and infinity.

The properties of the singlet state are investigated. Its ground-state energy
and the occupations $f_{0},f_{1},f_{2}$ of the d-states are in \ good
agreement with the extensive calculations by Gunnarsson and Schoenhammer using
the large $N_{f}$-expansion. However, while in the large $N_{f}$-expansion one
has to go to a large basis of states to obtain a good ground-state energy the
present solution is extremely compact.

The spectral composition of the two AFR states is quite interesting. Their
composition is quite different away from the Fermi energy. Close to the Fermi
energy one finds a large peak in the occupation density which saturates only
for very small energies. This low energy occupation is essentially identical
for the spin up and down AFR state.

The structure presented here of the ground state allows a number of
variations. Instead of using just two bases $\left\{  a_{0+}^{\ast}%
,a_{i+}^{\ast}\right\}  $ and $\left\{  a_{0-}^{\ast},a_{i-}^{\ast}\right\}  $
one can use four or eight bases, for each of the $\psi_{X}$ $\left(
X=A,B,C,D\right)  $ a different one for spin up and down. These solutions
improve the ground-state energy only slightly for the singlet state.

A detailed analysis of the present solution is planed. For example, the
construction of the triplet state and the calculation of transport scattering
by the impurity in this ground state are desirable. Above all, it is of
interest whether an extension of the presented solution can contribute to the
periodic Anderson impurity problem.

\newpage

\appendix

\section{ Some Details about the Numerical Calculations}

\subsection{ Wilson's s-electron basis}

Wilson \cite{W18} in his Kondo paper considers an s-band with energy values
ranging from $-1$ to 1. In the next step Wilson replaced the continuum of
s-states by a discrete set of states. This is done on a logarithmic scale. The
discrete energy values are 1, $1/\Lambda$, $1/\Lambda^{2}$, etc and $-1$,
$-1/\Lambda$, $-1/\Lambda^{2}$, etc where $\Lambda$ is a parameter larger than
one. (In this paper $\Lambda=2$ is chosen). These discrete $\xi_{\nu}$ points
are used to define a sequence of intervals: the interval $\nu$ (for $\nu$%
$<$%
N/2) is $\xi_{\nu-1}=-1/2^{\nu-1}<$ $\varepsilon$ $<-1/2^{\nu}=\xi_{\nu}$
(there are equivalent intervals for positive $\xi$-values where $\nu$ is
replace by $\left(  N-\nu\right)  $ but we discuss here only the negative
energies). The new Wilson states $c_{\nu}^{\ast}$ are a superposition of all
states in the energy interval $\left(  \xi_{\nu-1},\xi_{\nu}\right)  $ and
have an (averaged) energy $\left(  \xi_{\nu}+\xi_{\nu-1}\right)
/2=\allowbreak\left(  -\dfrac{3}{2}\right)  \dfrac{1}{2^{\nu}}$, i.e.
$-\frac{3}{4},-\frac{3}{8},-\frac{3}{16},..,-\frac{3}{2\cdot2^{N/2}},-\frac
{1}{2\cdot2^{N/2}}.$ This spectrum continues symmetrically for positive energies.

The amplitude of the states at the origin are chosen to be $\phi_{\nu}\left(
0\right)  =$ $\left(  \xi_{\nu}-\xi_{\nu-1}\right)  /\sqrt{2}=1/2^{\nu+1}$. A
state which is homogeneously composed of all energies in the full band has
than an amplitude of "1".

Therefore this choice of $s^{\ast}$-states yields a dependence of the s-d
matrix element $V_{sd}\left(  \nu\right)  $ on the state $\nu.$ The essential
advantage of the Wilson basis is that it has an arbitrarily fine energy
spacing at the Fermi energy.

\subsection{ Construction of the Basis $a_{0}^{*}$, $a_{i}^{*}$}

For the construction of the state $a_{0}^{\ast}$ and the rest of basis
$a_{i}^{\ast}$ one starts with the s-band electrons $\left\{  c_{\nu}^{\ast
}\right\}  $ which consists of $N$ states (for example Wilson's states). The
$d^{\ast}$-state is ignored for the moment. \newline

\begin{itemize}
\item In step (1) one forms an normalized state $a_{0}^{\ast}$ out of the
s-states with:
\end{itemize}

\begin{equation}
a_{0}^{\ast}=\sum_{\nu=1}^{N}\alpha_{\nu}^{0}c_{\nu}^{\ast}%
\end{equation}
The coefficients $\alpha_{\nu}^{0}$ can be at first arbitrary. One reasonable
choice is $\alpha_{\nu}^{0}=1/\sqrt{N}$

\begin{itemize}
\item In step (2) $\left(  N-1\right)  $ new basis states $a_{i}^{\ast}$
$\left(  1\leq i\leq N-1\right)  $ are formed which are normalized and
orthogonal to each other and to $a_{0}^{\ast}$.

\item In step (3) the s-band Hamiltonian $H_{0}$ is constructed in this new
basis. One puts the state $a_{0}^{\ast}$ at the top so that its matrix
elements are $H_{0i}$ and $H_{i0}$.

\item In step (4) the $\left(  N-1\right)  $-sub Hamiltonian which does not
contain the state $a_{0}^{\ast}$ is diagonalized. The resulting Hamilton
matrix for the s-band then has the form%
\begin{equation}
H_{0}=\left(
\begin{array}
[c]{ccccc}%
E(0) & V_{fr}(1) & V_{fr}(2) & ... & V_{fr}(N-1)\\
V_{fr}(1) & E(1) & 0 & ... & 0\\
V_{fr}(2) & 0 & E(2) & ... & 0\\
.. & ... & ... & ... & ...\\
V_{fr}(N-1) & 0 & 0 & ... & E(N-1)
\end{array}
\right)  \label{hmat}%
\end{equation}
The creation operators of the new basis are given by a new set of $\left\{
a_{i}^{\ast}\right\}  ,$ ($0<i\leq N-1)$. Again the $a_{i}^{\ast}$ can be
expressed in term of the s-states; $a_{i}^{\ast}=\sum_{\nu=1}^{N}\alpha_{\nu
}^{i}c_{\nu}^{\ast}$. After the state $a_{0}^{\ast}$ is constructed the other
states $a_{i}^{\ast}$ are uniquely determined. The additional s-d hopping
Hamiltonian can be expressed in the terms of new basis and one obtains the
Friedel Hamiltonian as given in eq. (\ref{hfr'}). The state $\Psi_{SS}$ is
formed and its energy expectation value is calculated.

\item In the final step (5) the state $a_{0}^{\ast\text{ }}$is rotated in the
$N$ dimensional Hilbert space until one reaches the absolute minimum of the
energy expectation value. In the example of the Friedel resonance Hamiltonian
this energy agrees numerically with an accuracy of $10^{-15}$ with the exact
ground-state energy of the Friedel Hamiltonian \cite{B91}.
\end{itemize}

\subsection{ The effective s-d matrix element for the multi-electron states}

The calculation of the energy expectation value requires the calculation of
many-electron matrix elements in different bases. We sketch here an example.
We consider the more general case that we have two wave functions $\Psi
_{A}=a_{0-}^{\ast}a_{0+}^{\ast}\prod_{i=1,\sigma}^{n-1}a_{i\sigma}^{\ast}%
\Phi_{0}$ and $\Psi_{B}=b_{0-}^{\ast}d_{+}^{\ast}\prod_{i=1,\sigma}%
^{n-1}b_{i\sigma}^{\ast}\Phi_{0}.$ Each is built from two different bases:
$\left\{  a_{0+},a_{i+}\right\}  ,$ $\left\{  a_{0-},a_{i-}\right\}  $ and
$\left\{  b_{0+},b_{i+}\right\}  ,\left\{  b_{0-},b_{i-}\right\}  $ (only
within this section the operators $b_{0,}^{\ast}b_{i}^{\ast}$ are used for the
AFR states to distinguish the different basis systems). The energy expectation
value contains for example a matrix element of the form $\langle\Psi
_{B}|H_{sd}^{+}|\Psi_{A}\rangle$. Here the s-d Hamiltonian $H_{sd}^{+}$ can be
expressed in any basis but for this matrix element the $a_{0+}^{\ast}$
representation is the optimal one. For the above matrix element one needs only
the hopping for spin up $\left(  +\right)  :$%
\begin{align}
H_{sd}^{+}  &  =\sum_{i=0}^{N-1}V_{sd}^{a+}\left(  i\right)  [d_{+}^{\ast
}a_{i+}+a_{i+}^{\ast}d_{+}]\label{hsd}\\
V_{sd}^{a+}(i)  &  =\sum_{\nu}V_{sd}\left(  \nu\right)  \alpha_{\nu+}%
^{i}\nonumber
\end{align}

The only term in $H_{sd}^{+}$ which yields a non-vanishing contribution to
$\langle V_{sd}^{AB}\rangle$ is $\langle\Psi_{B}|\sum_{i=0}^{N-1}V_{sd}%
^{a+}\left(  i\right)  d_{+}^{\ast}a_{i+}|$ $\Psi_{A}\rangle.$

This matrix element contains \newline(a) the multi-scalar product of the two
$n$ electron states for spin down $F^{AB}=\langle b_{0-}^{\ast}\prod
_{i=1}^{n-1}b_{i-}^{\ast}\Phi_{0}|a_{0-}^{\ast}\prod_{i=1}^{n-1}a_{i-}^{\ast
}\Phi_{0}\rangle$ and \newline(b) the matrix element $M_{sd}^{AB}=\langle
d_{+}^{\ast}\prod_{i=1}^{n-1}b_{i+}^{\ast}\Phi_{0}|\sum_{i=0}^{N-1}V_{sd}%
^{a+}\left(  i\right)  d_{+}^{\ast}a_{i+}|a_{0+}^{\ast}\prod_{i=1}^{n-1}%
a_{i+}^{\ast}\Phi_{0}\rangle.$ \newline The multi-scalar product is a
determinant of order $n$ containing the single electron scalar products
between all occupied states.%

\begin{equation}
F^{AB}=\left\vert
\begin{array}
[c]{cccc}%
\left\langle b_{0-}^{\ast}|a_{0-}^{\ast}\ \right\rangle  & \left\langle
b_{0-}^{\ast}|a_{1-}^{\ast}\ \right\rangle  & ... & \left\langle b_{0-}^{\ast
}|a_{\left(  n-1\right)  -}^{\ast}\ \right\rangle \\
\left\langle \ b_{1-}^{\ast}|a_{0-}^{\ast}\ \right\rangle  & \left\langle
\ b_{1-}^{\ast}|a_{1-}^{\ast}\right\rangle  & ... & \left\langle
\ b_{1-}^{\ast}|a_{\left(  n-1\right)  -}^{\ast}\right\rangle \\
... & ... & ... & ...\\
\left\langle \ \ b_{\left(  n-1\right)  -}^{\ast}|a_{0-}^{\ast}\right\rangle
& \left\langle b_{\left(  n-1\right)  -}^{\ast}|a_{1-}^{\ast}\ \right\rangle
& ... & \left\langle \ b_{\left(  n-1\right)  -}^{\ast}|a_{\left(  n-1\right)
-}^{\ast}\right\rangle
\end{array}
\right\vert \label{fab}%
\end{equation}

When the two AFR states are identical then the underlying matrix becomes the
unity matrix.

Part (b) yields%

\begin{equation}
\langle M_{sd}^{AB}\rangle=\left\vert
\begin{array}
[c]{cccc}%
V_{sd}^{a+}(0) & V_{sd}^{a+}(1) & ... & V_{sd}^{a+}\left(  n-1\right) \\
\left\langle \ b_{1+}^{\ast}|a_{0+}^{\ast}\right\rangle  & \left\langle
\ b_{1+}^{\ast}|a_{1+}^{\ast}\right\rangle  & ... & \left\langle
\ b_{1+}^{\ast}|a_{\left(  n-1\right)  +}^{\ast}\right\rangle \\
... & ... & ... & ...\\
\left\langle b_{\left(  n-1\right)  +}^{\ast}|a_{0+}^{\ast}\ \right\rangle  &
\left\langle \ b_{\left(  n-1\right)  +}^{\ast}|a_{1+}^{\ast}\right\rangle  &
... & \left\langle b_{\left(  n-1\right)  +}^{\ast}|a_{\left(  n-1\right)
+}^{\ast}\right\rangle
\end{array}
\right\vert \label{vab}%
\end{equation}

\subsection{Details of the comparison with Gunnarsson and Schoenhammer's
numerical evaluation}

Gunnarsson and Schoenhammer (GS) \cite{G34} applied the large $N_{f}$ method
to finite Coulomb interaction and spin $1/2$. They calculated the ground-state
energy in different approximations. Since it is interesting to compare their
results with the present calculation I sketch briefly the different states
they considered. The corresponding graphical sketch of these states can be
found in figure 1 of ref. \cite{G34}. These states are collected in table
\ref{T2}. The first column gives GS's code for the state, the second column
shows in which power of the $1/N_{f}$ expansion the state occurs, the third
column gives the occupation of the d-level in the considered state, the fourth
column the number of holes and electrons (above the Fermi energy) in the
s-band, and finally the fifth column gives the number of amplitudes
(parameters) which one has to optimize in the numerical evaluation (again $N$
is the number of band states in the numerical evaluation). As an example the
state \textquotedblright d\textquotedblright\ is part of the $\left(
1N_{f}\right)  ^{1}$ expansion, it has, for example, the d$\uparrow$-state
occupied, the s-band has one hole in the s$\uparrow$-band, another
electron-hole pair is either in the s$\uparrow$- or s$\downarrow$-band. The
total multiplicity of the state \textquotedblright d\textquotedblright\ is
therefore $2\ast\left(  N/2\right)  \ast2\left(  N/2\right)  ^{2}$. The
prefactor 2*2 is replaced by \textquotedblright$\alpha"$ in column 5
($\alpha\geq1$).%

\begin{align*}
&
\begin{tabular}
[c]{|l|l|l|l|l|}\hline
\textbf{%
\begin{tabular}
[c]{l}%
name\\
of\\
state
\end{tabular}
} & \textbf{%
\begin{tabular}
[c]{l}%
power\\
of\\
expans.
\end{tabular}
} & \textbf{%
\begin{tabular}
[c]{l}%
occup. of\\
d-states
\end{tabular}
} &
\begin{tabular}
[c]{lllll}
&  &  & \textbf{s-band} &
\end{tabular}
& \textbf{%
\begin{tabular}
[c]{l}%
no of\\
param.
\end{tabular}
}\\\hline%
$\vert$%
0$\rangle$ & $\left(  1/N_{f}\right)  ^{0}$ & empty d-states & half occupied
band for spin $\uparrow$ and $\downarrow$ & 0\\\hline
a & $\left(  1/N_{f}\right)  ^{0}$ & 1 d$\uparrow$ or d$\downarrow$ & one
$\uparrow$- or $\downarrow$-hole in
$\vert$%
0$\rangle$ & $N/2$\\\hline
b & $\left(  1/N_{f}\right)  ^{0}$ & 2 d-states & $\uparrow$-hole and
$\downarrow$-hole in
$\vert$%
0$\rangle$ & $\alpha\left(  N/2\right)  ^{2}$\\\hline
c & $\left(  1/N_{f}\right)  ^{1}$ & empty d-states & one $\uparrow$- or
$\downarrow$ electron-hole pair in
$\vert$%
0$\rangle$ & $\alpha\left(  N/2\right)  ^{2}$\\\hline
d & $\left(  1/N_{f}\right)  ^{1}$ & 1 d$\uparrow$ or d$\downarrow$ & two
holes in
$\vert$%
0$\rangle$ and one electron & $\alpha\left(  N/2\right)  ^{3}$\\\hline
e & $\left(  1/N_{f}\right)  ^{1}$ & 2 d-states & three holes in
$\vert$%
0$\rangle$ and one electron & $\alpha\left(  N/2\right)  ^{4}$\\\hline
f & $\left(  1/N_{f}\right)  ^{2}$ & empty d-states & two $\uparrow$- or
$\downarrow$ electron-hole pairs in
$\vert$%
0$\rangle$ & $\alpha\left(  N/2\right)  ^{4}$\\\hline
g & $\left(  1/N_{f}\right)  ^{2}$ & 1 d$\uparrow$ or d$\downarrow$ & three
holes in
$\vert$%
0$\rangle$ and two electrons & $\alpha\left(  N/2\right)  ^{5}$\\\hline
h & $\left(  1/N_{f}\right)  ^{2}$ & 2 d-states & four holes in
$\vert$%
0$\rangle$ and two electrons & $\alpha\left(  N/2\right)  ^{6}$\\\hline
\end{tabular}
\\
&
\begin{tabular}
[c]{l}%
Table 3: Gunnarsson and Schoenhammer states
\end{tabular}
\end{align*}

G\&S use a (different) exponential energy mesh of the form $\varepsilon
_{i}=\pm\left[  \alpha-\exp\left(  x_{i}\right)  \right]  $. They use the
value $\alpha=0.2,$ and $x_{i}$ lies in the range $\left(  \ln\left(
\alpha\right)  ,\ln\left(  \alpha+B_{GS}\right)  \right)  $ $=\left(
-1.\,\allowbreak609\,4,1.\,\allowbreak824\,5\right)  .$This means $x_{i}$
takes the values $x_{i}=\ln\left(  \alpha\right)  +i/N\ast\left[  \ln\left(
\alpha+B_{GS}\right)  -\ln\left(  \alpha\right)  \right]  $. G\&S used for $N$
the values 9, 19, 29 and extrapolated to $N\rightarrow\infty$. For this
comparison I use the corresponding energy mesh and extrapolation. The only
difference is that in my calculation the $\varepsilon_{i}$ yield the energy
frame and the energy states lie in the center between two $\varepsilon_{i}$
whereas G\&S used the $\varepsilon_{i}$ as their energy states. After the
extrapolation towards $N\rightarrow\infty$ this difference should be negligible.

The energy dependent s-d-matrix element $V\left(  \varepsilon\right)  $ adds a
complication in the numerical evaluation. It varies strongly with energy. I
average $\left[  V\left(  \varepsilon\right)  \right]  ^{2}$ over each energy
range.

\end{document}